\documentclass[superscriptaddress,pra,twocolumn,nofootinbib,notitlepage]{revtex4-1}
\usepackage{graphicx}
\usepackage{amsmath}
\usepackage{amssymb}
\usepackage{amsthm}
\usepackage{amsfonts}
\usepackage{comment}
\usepackage[colorlinks]{hyperref}
\usepackage[caption=false]{subfig}
\usepackage[all]{hypcap}
\usepackage{tikz}
\usepackage{verbatim}
\usepackage{subfig}
\usetikzlibrary{arrows}
\usepackage{units}
\usepackage{soul}
\usepackage{algorithm}
\usepackage{algorithmic}
\usepackage{url}

\usetikzlibrary{external}
\newcommand{\eq}[1]{Eq.~\hyperref[eq:#1]{(\ref*{eq:#1})}}
\renewcommand{\sec}[1]{\hyperref[sec:#1]{Section~\ref*{sec:#1}}}
\DeclareRobustCommand{\app}[1]{\hyperref[app:#1]{Appendix~\ref*{app:#1}}}
\newcommand{\tab}[1]{\hyperref[tab:#1]{Table~\ref*{tab:#1}}}
\newcommand{\fig}[1]{\hyperref[fig:#1]{Figure~\ref*{fig:#1}}}
\newcommand{\figa}[2]{\hyperref[fig:#1]{Figure~\ref*{fig:#1}#2}}
\newcommand{\figx}[2]{\hyperref[fig:#1]{Figure~\ref*{fig:#1}(#2)}}
\newcommand{\thm}[1]{\hyperref[thm:#1]{Theorem~\ref*{thm:#1}}}
\newcommand{\lem}[1]{\hyperref[lem:#1]{Lemma~\ref*{lem:#1}}}
\newcommand{\cor}[1]{\hyperref[cor:#1]{Corollary~\ref*{cor:#1}}}
\newcommand{\defn}[1]{\hyperref[def:#1]{Definition~\ref*{def:#1}}}
\newcommand{\alg}[1]{\hyperref[alg:#1]{Algorithm~\ref*{alg:#1}}}
\newcommand{\clm}[1]{\hyperref[claim:#1]{Claim~\ref*{claim:#1}}}

\def\avg#1{\mathinner{\langle{#1}\rangle}}

\newcommand{\proj}[1]{\mathinner{|{#1}\rangle}\!\!\mathinner{\langle{#1}|}}

\input{Qcircuit}

\begin{document}

\title{Quantum Simulation of the Sachdev-Ye-Kitaev Model by Asymmetric Qubitization}

\date{\today}
\author{Ryan Babbush}
\email[Corresponding author: ]{babbush@google.com}
\affiliation{Google Inc., Venice, CA 90291, United States}
\author{Dominic W. Berry}
\affiliation{Macquarie University, Sydney, NSW 2109, Australia}
\author{Hartmut Neven}
\affiliation{Google Inc., Venice, CA 90291, United States}

\begin{abstract}
We show that one can quantum simulate the dynamics of a Sachdev-Ye-Kitaev model with $N$ Majorana modes for time $t$ to precision $\epsilon$ with gate complexity ${\cal O}(N^{7/2} t + N^{5/2} t \,{\rm polylog}(N/ \epsilon))$. In addition to scaling sublinearly in the number of Hamiltonian terms, this gate complexity represents an exponential improvement in $1/\epsilon$ and large polynomial improvement in $N$ and $t$ over prior state-of-the-art algorithms which scale as ${\cal O}(N^{10} t^2 / \epsilon)$. Our approach involves a variant of the qubitization technique in which we encode the Hamiltonian $H$ as an asymmetric projection of a signal oracle $U$ onto two different signal states prepared by state oracles, $A\left\vert{0}\right\rangle \mapsto \left\vert{A}\right\rangle$ and $B \left\vert{0}\right\rangle \mapsto \left\vert{B}\right\rangle$, such that $H = \left\langle{B}\right\vert U\left\vert{A}\right\rangle$. Our strategy for applying this method to the Sachdev-Ye-Kitaev model involves realizing $B$ using only Hadamard gates and realizing $A$ as a random quantum circuit.
\end{abstract}

\maketitle

\section{Introduction}

The AdS/CFT correspondence is a conjectured relationship between the quantum physics of correlated many-body systems and the classical physics of gravity in one higher dimension \cite{Maldacena1999}. Holographic dualities such as AdS/CFT have become increasingly important tools for studying quantum gravity; however, it has generally been difficult to find simple models that capture exotic features such as black holes \cite{Maldacena2016}. Introduced by Kitaev \cite{Kitaev2015} based on a variant of an earlier model by Sachdev and Ye \cite{Sachdev1993}, the Sachdev-Ye-Kitaev (SYK) model has been widely studied in recent years as an example of a simple quantum many-body system which may have an interesting holographic dual. The SYK model can be expressed as
\begin{equation}
\label{eq:syk}
H = \frac{1}{4\cdot 4!} \sum_{p,q,r,s=0}^{N-1} J_{pqrs} \gamma_p \gamma_q \gamma_r \gamma_s
\end{equation}
where the $J_{pqrs}$ are real-valued scalars drawn randomly from a normal distribution with variance $\sigma^2 = 3!J^2/N^3$ and the $\gamma_p$ are Majorana fermion mode operators.

A feature of the SYK model is that for large $N$ and strong coupling $J$ it is possible to sum all of the Feynman diagrams to obtain (among other properties) out-of-order-time correlation functions \cite{Kitaev2015}. Such methods reveal that the SYK model is maximally chaotic (having the same Lyapunov exponent as black holes in Einstein gravity) \cite{Kitaev2015}. This property, in conjunction with the emergence of an approximate conformal symmetry at low temperatures, suggests that the holographic dual of the SYK model is a theory of Einstein gravity \cite{Polchinski2016}.

However, questions remain about the SYK model which have not yet been solved by analytical methods. Since quantum computers are widely conjectured to be capable of modeling otherwise intractable quantum systems \cite{Feynman1982}, they may also prove useful tools for studying the quantum many-body side of such holographic duals. For example, it has been analytically challenging to obtain the density of states associated with interesting states (e.g.\ the thermal state) of the SYK model. A quantum computer could efficiently sample such distributions by performing quantum phase estimation \cite{Kitaev1995}.

Earlier work \cite{Garcia-Alvarez2017} has proposed a scheme for quantum simulating SYK model dynamics using a Lie-Trotter based algorithm \cite{Trotter1959} with gate complexity ${\cal O}(N^{10} t^2 / \epsilon)$, where $t$ is time and $\epsilon$ is target precision. Yet such scaling suggests that interesting (e.g.\ $N > 100$) SYK model quantum simulations would remain intractable for even a fault-tolerant quantum computer.
In this paper we describe an algorithm to quantum simulate the SYK model with gate complexity ${\cal O}(N^{7/2} t + N^{5/2} t \, {\rm polylog}(N / \epsilon))$.
We compile the bottleneck components of our approach to Clifford + T gates and find that interesting simulations are possible with fewer than ten million $Jt$ T gates. This relatively low T complexity makes the application attractive because T gates consume many logical qubits and take much longer to apply than any other operation within all known two-dimensional error-correcting codes with low thresholds (e.g.\ the surface code) \cite{Fowler2012}.

\section{Asymmetric qubitization}

Our method is to use a linear combinations of unitaries (LCU) approach combined with an ``asymmetric'' extension of the qubitization simulation framework \cite{Low2016}.
The original formulation of qubitization (we refer to as ``symmetric'' qubitization) requires a state preparation with amplitudes proportional to the \emph{square roots} of the coefficients $J_{pqrs}$.
Instead, our asymmetric form of qubitization permits a state preparation with amplitudes \emph{linearly} proportional to the normally distributed $J_{pqrs}$.
Such a state can be prepared with low gate complexity by using a random quantum circuit.

The standard LCU query model \cite{Childs2012} for Hamiltonian simulation uses two unspecified unitary operators, $G$ and $U$. The operator $G$ initializes a state, $G \ket{0} \mapsto \ket{G}$ in an ancilla register.
Then $U$ performs one of a set of unitaries on the target system controlled on the ancilla register.
Finally, measurement of this ancilla register in the state $\ket{G}$ gives the linear combination of unitaries applied to the target system. Conventional qubitization \cite{Low2016} requires that
\begin{equation}
\label{eq:signal}
\bra{G} U \ket{G}  = \frac{H}{\lambda},
\end{equation}
which then enables Hamiltonian simulation to be achieved via quantum signal processing \cite{Low2017}. 
In contrast to the LCU model, no postselection or amplitude amplification is required for quantum signal processing.
In \eq{signal}, $U$ acts on the system and ancilla registers whereas $G$ is a state only on the ancilla register. Therefore $\bra{G} U \ket{G}$ gives an operator acting on the target system. We have included a scaling factor $\lambda$, since usually an operator proportional to the desired Hamiltonian is given. For our application, $U$ is a controlled unitary operation between the ancilla and target systems, but the qubitization framework allows $U$ to be more general.

For our SYK model simulation we would like to encode the Hamiltonian using two different state preparation oracles; i.e., in contrast to \eq{signal}, we would like to have state preparations $A \ket{0} \mapsto \ket{A}$ and $B \ket{0} \mapsto \ket{B}$ together with a unitary $V$ such that $\bra{B} V \ket{A} = H/\lambda$. This is what we mean by ``asymmetric'' qubitization. In order to use qubitization with these asymmetric states, we can bundle the operations $A$ and $B$ together with $V$, so $\bra{0}B^\dagger VA\ket{0}=H/\lambda$.
The difficulty now is that $B^\dagger VA$ will not be self-inverse, but qubitization works best with self-inverse operators.

We will take $V$ to be self-inverse. That means we can add an ancilla qubit and construct a self-inverse $U$ as (in a block matrix representation)
\begin{equation}
U = \begin{bmatrix}
0 & A^\dagger VB \\
B^\dagger VA & 0
\end{bmatrix} = \begin{bmatrix}
A^\dagger & 0 \\
0 & B^\dagger
\end{bmatrix}\begin{bmatrix}
0 & V \\
V & 0
\end{bmatrix}\begin{bmatrix}
A & 0 \\
0 & B
\end{bmatrix}.
\end{equation}
That is, the ancilla controls which state preparation is performed, then we perform $V$ and a \textsc{not} on the ancilla qubit, then there is an inverse controlled state preparation.
The operation $V$ is the most costly and is only performed once.
The complexity of $A$ and $B$ is logarithmic compared to the complexity of $V$, so the overall complexity is not significantly increased.
Reference \cite{Low2016} proposes a method to construct self-inverse operators, but it doubles the complexity.
An alternative realization of asymmetric qubitization is discussed in the \app{oaa}.

In our case, we take the Hamiltonian to be a sum of self-inverse terms
\begin{equation}
\label{eq:Hsum}
H = \sum_{\ell=0}^{L-1} w_\ell H_\ell .
\end{equation}
and $V$ will be a controlled $H_\ell$,
\begin{equation}
\label{eq:select}
V \equiv \sum_{\ell=0}^{L-1} \proj{\ell} \otimes H_\ell .
\end{equation}

The state preparation $A$ will be a random orthogonal operation producing Gaussian distributed amplitudes, and $B$ will just give an equal superposition.

To perform Hamiltonian simulation with asymmetric qubitization, we can use the same method as for symmetric qubitization, with the new operation $U$ constructed above, and $G$ just being a Hadamard on the new ancilla qubit.
For quantum signal processing \cite{Low2017}, one combines the operation $U$ with a reflection $R=2\proj{G}-\openone$ as ${\cal W}=R U$.
Applications of ${\cal W}$ and ${\cal W}^\dagger$ are controlled by a qubit, and these controlled-${\cal W}$ operations are interspersed with rotations of the control qubit.
Using $M$ controlled operations can yield an overall polynomial in ${\cal W}$ of order $M$ \cite{Low2017}.
This polynomial is chosen to give the Hamiltonian evolution $e^{-iHt}$.
The order of the polynomial required scales as $\lambda$, which therefore governs the complexity.
In the following we will examine the additional cost in $\lambda$ due to the asymmetric qubitization, explain how to perform the state preparations and controlled operation $V$, then determine the overall cost. 

Suppose that operators $A$ and $B$ prepare states $\ket{A}$ and $\ket{B}$ as
\begin{align}
\label{eq:a}
\ket{A} & \equiv A \ket{0}^{\otimes \log L} = \sum_{\ell=0}^{L-1} \alpha_\ell \ket{\ell}\\
\ket{B} & \equiv B \ket{0}^{\otimes \log L} = \sum_{\ell=0}^{L-1} \beta_\ell \ket{\ell}.
\end{align}
Then, taking the register $\ket{\psi}$ to be the target subsystem on which the operators $H_\ell$ act,
\begin{align}
& \frac{H} {\lambda}\ket{\psi} = \bra{0} B^\dagger U A\ket{0}\ket{\psi}
= \sum_{\ell=0}^{L-1} \alpha_\ell \beta_\ell^* H_\ell \ket{\psi}.
\end{align}
Thus, using \eq{Hsum} we obtain,
\begin{equation}
H = \lambda \sum_{\ell=0}^{L-1} \alpha_\ell \beta_\ell^* H_\ell = \sum_{\ell=0}^{L-1} w_\ell H_\ell
\end{equation}
so that for all $\ell$, we require $\lambda \alpha_\ell \beta_\ell^* = w_\ell$. Therefore, taking the absolute value and summing gives
\begin{equation}
\label{eq:lamform}
\lambda \sum_{\ell=0}^{L-1} |\alpha_\ell\beta_\ell^*|=\sum_{\ell=0}^{L-1} |w_\ell| \leq \lambda,
\end{equation}
which comes from the Cauchy-Schwarz inequality applied to normalized states $\ket{A}$ and $\ket{B}$. Hence, $\lambda$ takes its minimum value of $\sum_\ell |w_\ell|$ in the case that $|\alpha_\ell|=|\beta_\ell|$.
This means that the most efficient case is that where the preparation and inverse preparation are symmetric.

For some classes of Hamiltonians, the preparation can be made heavily asymmetric without much cost.
In particular, consider the asymmetric case where
\begin{equation}
\alpha_\ell = \frac{w_\ell}{\sqrt{\sum_{\ell=0}^{L-1} |w_\ell|^2}},\qquad \beta_\ell = \frac 1{\sqrt{L}}.
\end{equation}
Then using \eq{lamform} gives
\begin{align}
\lambda \sum_{\ell=0}^{L-1} \frac{|w_\ell|}{\sqrt{\sum_{\ell=0}^{L-1} |w_\ell|^2}}\frac{1}{\sqrt{L}} = \sum_{\ell=0}^{L-1} |w_\ell|, \\
\lambda = \sqrt{L\sum_{\ell=0}^{L-1} |w_\ell|^2} = L \sqrt{\avg{\left |w_\ell\right |^2}} .
\label{eq:gaussian_lambda}
\end{align}
Using $\langle \cdot \rangle$ to indicate the mean over $\ell$, this means that the overhead from using the asymmetric $\ket{A}$ and $\ket{B}$ over the symmetric case is $\sqrt{\avg{|w_\ell|^2}} / \avg{|w_\ell|}$.
In the case where the $w_\ell$ are drawn from a normal distribution (e.g.\ the SYK model), the additional complexity is only $\sqrt{\pi/2}\approx 1.25$.

\section{SYK Model LCU Oracles}
\subsection{The State Preparation Circuits}

Our strategy for implementing $A$ and $B$ oracles to simulate the SYK model is straightforward; $B$ will consist of Hadamard gates which initialize the symmetric superposition state and $A$ will be a random quantum circuit with orthogonal rotations. Let us assume for simplicity that the number of terms in our SYK model Hamiltonian ($L = N^4$) is a binary power. Then, one can initialize the state $\ket{B}$ such that $\beta_\ell = 1 / \sqrt{L}$ by implementing the circuit $B$ as a sequence of $\log L$ Hadamard gates.

Consider the state $\ket{A}$ output by a random quantum circuit $A$ with orthogonal rotations. We will use orthogonal rotations in order to ensure that the amplitudes remain real. Using the definition in \eq{a}, known properties of orthogonal random quantum evolutions \cite{PorterThomas} hold that the $\alpha_\ell$ are Gaussian distributed with zero mean and variance equal to the Hilbert space dimension. This approach to simulating the SYK model reveals a possibly interesting connection between chaos in random quantum circuits \cite{Boixo2016} and chaos in $\textrm{AdS}_2$ holography \cite{Jensen2016}. 

The asymmetric state preparation requires that the $\alpha_\ell$ are proportional to the desired weightings for the terms in the Hamiltonian. Because they have a normal distribution, they correctly generate the weights $w_{\ell}$. There is a difference in the variance, but the variance only affects an overall scaling of the values of the $\alpha_\ell$, which is what we expect because the $\alpha_\ell$ correspond to the amplitudes of a normalized state.
The scaling factor is taken into account in $\lambda$, giving the complexity of the calculation.

The value of $\lambda$ for this asymmetric state preparation is given by \eq{gaussian_lambda}, which only depends on $w_\ell$ and the dimension.
It does not explicitly depend on $\alpha_\ell$, because that proportionality is already taken into account by the fact that the $\alpha_\ell$ are normalized.
For the SYK model,
\begin{equation}
w_{pqrs} \equiv \frac 1{4\cdot 4!} J_{pqrs},
\end{equation}
where we have changed the index to $pqrs$ to match the notation used for $J$.
Then, the mean square value is
\begin{align}
\langle |w_{pqrs}|^2\rangle = \frac 1{(4\cdot 4!)^2} \langle J_{pqrs}^2\rangle
\approx  \frac 1{(4\cdot 4!)^2} \frac{3!}{N^3} J^2.
\end{align}
There is an approximate equality here, because the mean of $w_{pqrs}$ indicates the mean summing over $pqrs$, not the expectation value of the probability distribution according to which $J_{pqrs}$ are chosen.
Using \eq{gaussian_lambda} then gives
\begin{equation}
\lambda \approx L \sqrt{\frac 1{(4\cdot 4!)^2} \frac{3!}{N^3} J^2} = N^{5/2} J \frac{\sqrt{3!}}{4\cdot 4!}.
\end{equation}

An outstanding question is how large the orthogonal random quantum circuits should be in order to achieve sufficient convergence to the Gaussian distribution in the coefficients. There have been many theoretical results on related questions such as the convergence of random quantum circuits to $t$-designs of the Haar measure \cite{Harrow2009,Znidaric2008,Brown2015,Harrow2018}. For a one dimensional random quantum circuit on $\log L$ qubits, the circuit approaches an $\epsilon$-approximate 2-design in gate complexity ${\cal O}(\log^2 L + \log (L / \epsilon))$ \cite{Harrow2009}. However, for circuits in higher dimension the gate complexity to achieve similar states is closer to ${\cal O}(\log (L / \epsilon))$ \cite{Harrow2018}.

While these works typically do not focus on the convergence of amplitudes to a Gaussian distribution, this topic was recently studied numerically in \cite{Boixo2016}. There, authors found rapid convergence of the probabilities (squared amplitudes) to the Porter-Thomas distribution \cite{PorterThomas}, which corresponds to convergence of real and imaginary components of the amplitudes to a Gaussian distribution. Thus, to avoid an in depth discussion of the requisite circuit size, we will conservatively assume that to achieve amplitudes that are within $\epsilon$ distance of Gaussian distributed amplitudes, it suffices to use circuits of size ${\cal O}({\rm polylog}(L/\epsilon)) = {\cal O}({\rm polylog}(N/\epsilon))$.

\begin{figure*}[t]
\begin{minipage}[t]{.55\textwidth}
\centering
\begin{flushright}{
\resizebox{\linewidth}{!}{
\Qcircuit @R=0.7em @C=0.5em {
\\
\lstick{\ell_2}             &\qw      &\ctrlo{3} &\qw       &\qw       &\qw     &\qw      &\qw      &\qw      &\qw      &\qw       &\qw     &\qw      &\qw      &\qw      &\qw      &\qw      &\qw       &\qw       &\qw     &\qw      &\qw      &\qw      &\qw      &\qw       &\qw     &\qw      &\qw      &\qw      &\qw      &\qw      &\ctrl{3} &\qw &\\
\lstick{\ell_1}             &\qw      &\qw       &\ctrlo{3} &\qw       &\qw     &\qw      &\qw      &\qw      &\qw      &\qw       &\qw     &\qw      &\qw      &\qw      &\ctrl{3} &\qw      &\ctrlo{3} &\qw       &\qw     &\qw      &\qw      &\qw      &\qw      &\qw       &\qw     &\qw      &\qw      &\qw      &\qw      &\ctrl{3} &\qw      &\qw &\\
\lstick{\ell_0}             &\qw      &\qw       &\qw       &\ctrlo{3} &\qw     &\qw      &\qw      &\ctrl{3} &\qw      &\ctrlo{3} &\qw     &\qw      &\qw      &\ctrl{3} &\qw      &\qw      &\qw       &\ctrlo{3} &\qw     &\qw      &\qw      &\ctrl{3} &\qw      &\ctrlo{3} &\qw     &\qw      &\qw      &\qw      &\ctrl{3} &\qw      &\qw      &\qw &\\
\lstick{\text{control}}     &\ctrl{4} &\ctrl{1}  &\qw       &\qw       &\qw     &\qw      &\qw      &\qw      &\qw      &\qw       &\qw     &\qw      &\qw      &\qw      &\qw      &\ctrl{1} &\qw       &\qw       &\qw     &\qw      &\qw      &\qw      &\qw      &\qw       &\qw     &\qw      &\qw      &\qw      &\qw      &\qw      &\ctrl{1} &\qw &\\
       &         &          &\ctrl{1}  &\qw       &\qw     &\qw      &\qw      &\qw      &\ctrl{1} &\qw       &\qw     &\qw      &\qw      &\qw      &\ctrl{1} &\targ    &\ctrl{1}  &\qw       &\qw     &\qw      &\qw      &\qw      &\ctrl{1} &\qw       &\qw     &\qw      &\qw      &\qw      &\qw      &\ctrl{1} &\qw      &    &\\
       &         &          &          &\ctrl{1}  &\qw     &\ctrl{1} &\qw      &\ctrl{1} &\targ    &\ctrl{1}  &\qw     &\ctrl{1} &\qw      &\ctrl{1} &\qw      &         &          &\ctrl{1}  &\qw     &\ctrl{1} &\qw      &\ctrl{1} &\targ    &\ctrl{1}  &\qw     &\ctrl{1} &\qw      &\qw      &\ctrl{1} &\qw      &         &    &\\
       &         &          &          &          &\ctrl{2}&\targ    &\ctrl{3} &\qw      &         &          &\ctrl{4}&\targ    &\ctrl{5} &\qw      &         &         &          &          &\ctrl{6}&\targ    &\ctrl{7} &\qw      &         &          &\ctrl{8}&\targ    &\ctrl{1} &\ctrl{9} &\qw      &         &         &    &\\
 &         &\qw       &\qw       &\qw       &\targ   &\ctrl{1} &\targ    &\qw      &\ctrl{2} &\qw       &\targ   &\ctrl{3} &\targ    &\qw      &\qw      &\ctrl{4} &\qw       &\qw       &\targ   &\ctrl{5} &\targ    &\qw      &\ctrl{6} &\qw       &\targ   &\ctrl{7} &\qw      &         &         &         &         &    &\\
\lstick{\ket{\psi}_{000}}   &\qw      &\qw       &\qw       &\qw       &\gate{X}&\gate{Z} &\qw      &\qw      &\qw      &\qw       &\qw     &\qw      &\qw      &\qw      &\qw      &\qw      &\qw       &\qw       &\qw     &\qw      &\qw      &\qw      &\qw      &\qw       &\qw     &\qw      &\qw      &\qw      &\qw      &\qw      &\qw      &\qw &\\
\lstick{\ket{\psi}_{001}}   &\qw      &\qw       &\qw       &\qw       &\qw     &\qw      &\gate{X} &\qw      &\gate{Z} &\qw       &\qw     &\qw      &\qw      &\qw      &\qw      &\qw      &\qw       &\qw       &\qw     &\qw      &\qw      &\qw      &\qw      &\qw       &\qw     &\qw      &\qw      &\qw      &\qw      &\qw      &\qw      &\qw &\\
\lstick{\ket{\psi}_{010}}   &\qw      &\qw       &\qw       &\qw       &\qw     &\qw      &\qw      &\qw      &\qw      &\qw       &\gate{X}&\gate{Z} &\qw      &\qw      &\qw      &\qw      &\qw       &\qw       &\qw     &\qw      &\qw      &\qw      &\qw      &\qw       &\qw     &\qw      &\qw      &\qw      &\qw      &\qw      &\qw      &\qw &\\
\lstick{\ket{\psi}_{011}}   &\qw      &\qw       &\qw       &\qw       &\qw     &\qw      &\qw      &\qw      &\qw      &\qw       &\qw     &\qw      &\gate{X} &\qw      &\qw      &\gate{Z} &\qw       &\qw       &\qw     &\qw      &\qw      &\qw      &\qw      &\qw       &\qw     &\qw      &\qw      &\qw      &\qw      &\qw      &\qw      &\qw &\\
\lstick{\ket{\psi}_{100}}   &\qw      &\qw       &\qw       &\qw       &\qw     &\qw      &\qw      &\qw      &\qw      &\qw       &\qw     &\qw      &\qw      &\qw      &\qw      &\qw      &\qw       &\qw       &\gate{X}&\gate{Z} &\qw      &\qw      &\qw      &\qw       &\qw     &\qw      &\qw      &\qw      &\qw      &\qw      &\qw      &\qw &\\
\lstick{\ket{\psi}_{101}}   &\qw      &\qw       &\qw       &\qw       &\qw     &\qw      &\qw      &\qw      &\qw      &\qw       &\qw     &\qw      &\qw      &\qw      &\qw      &\qw      &\qw       &\qw       &\qw     &\qw      &\gate{X} &\qw      &\gate{Z} &\qw       &\qw     &\qw      &\qw      &\qw      &\qw      &\qw      &\qw      &\qw &\\
\lstick{\ket{\psi}_{110}}   &\qw      &\qw       &\qw       &\qw       &\qw     &\qw      &\qw      &\qw      &\qw      &\qw       &\qw     &\qw      &\qw      &\qw      &\qw      &\qw      &\qw       &\qw       &\qw     &\qw      &\qw      &\qw      &\qw      &\qw       &\gate{X}&\gate{Z} &\qw      &\qw      &\qw      &\qw      &\qw      &\qw &\\
\lstick{\ket{\psi}_{111}}   &\qw      &\qw       &\qw       &\qw       &\qw     &\qw      &\qw      &\qw      &\qw      &\qw       &\qw     &\qw      &\qw      &\qw      &\qw      &\qw      &\qw       &\qw       &\qw     &\qw      &\qw      &\qw      &\qw      &\qw       &\qw     &\qw      &\qw      &\gate{X} &\qw      &\qw      &\qw      &\qw &\\
\\
}
}}\end{flushright}
  \end{minipage}
\begin{minipage}[t]{.4\textwidth}
\centering
\centerline{
  \resizebox{\linewidth}{!}{
\Qcircuit @R=1em @C=0.75em {
\\
&\\
&\\
&\\
&\\
&\\
&\\
&\\
&\\
&\\
&\\
&\ctrl{1}  &\qw && &&&\qw                &\qw                &\ctrl{1}           &\qw && &&&\qw                &\qw      &\qw              &\ctrl{2} &\qw      &\qw      &\qw              &\qw      &\qw              &\qw \\
&\ctrl{1}  &\qw &&=&&&\qw                &\qw                &\ctrl{1}           &\qw &&=&&&\qw                &\ctrl{1} &\qw              &\qw      &\qw      &\ctrl{1} &\qw              &\qw      &\qw              &\qw \\
&          &\qw && &&&                   &\lstick{|0\rangle} &\targ              &\qw && &&&\lstick{|T\rangle} &\targ    &\gate{T^\dagger} &\targ    &\gate{T} &\targ    &\gate{T^\dagger} &\gate{H} &\gate{S^\dagger} &\qw \\
\\
&\ctrl{1}  &\qw && &&&\qw                &\ctrl{1}           &\qw                &\qw && &&&&\qw                &\qw      &\ctrl{1}          &\qw\\
&\ctrl{1}  &\qw &&=&&&\qw                &\ctrl{1}           &\qw                &\qw &&=&&&&\qw                &\qw      &\gate{Z}          &\qw\\
&\qw       &    && &&&\qw                &\targ              &\rstick{|0\rangle} \qw &&& &&&&\gate{H}           &\meter   &\control \cw \cwx &   \\
}
  }}
\label{fig:majorana}
\end{minipage}
  \caption{(Left): circuit to control the application of a Majorana operator as in \eq{select_majorana}. This circuit has a T-count of $4N-4$, where $N$ is the size of the target register. The circuit acts on $N + 2 \log N + 1$ qubits, including $\log N$ ancillae. Note that this diagram uses notation for setting an ancilla to the logical \textsc{and} of two other qubits, described in the figure to the right. Our implementation of $U$ consists of four of these circuits, which would require a total of $16 N-16$ T gates and $\log N$ ancillae. This circuit is explained in detail in Section 3B of \cite{BabbushSpectra}. (Right): circuit for computing and uncomputing \textsc{and} operations \cite{GidneyAdder}, defined in terms of Toffoli gates and Clifford+T gates.}
\end{figure*}
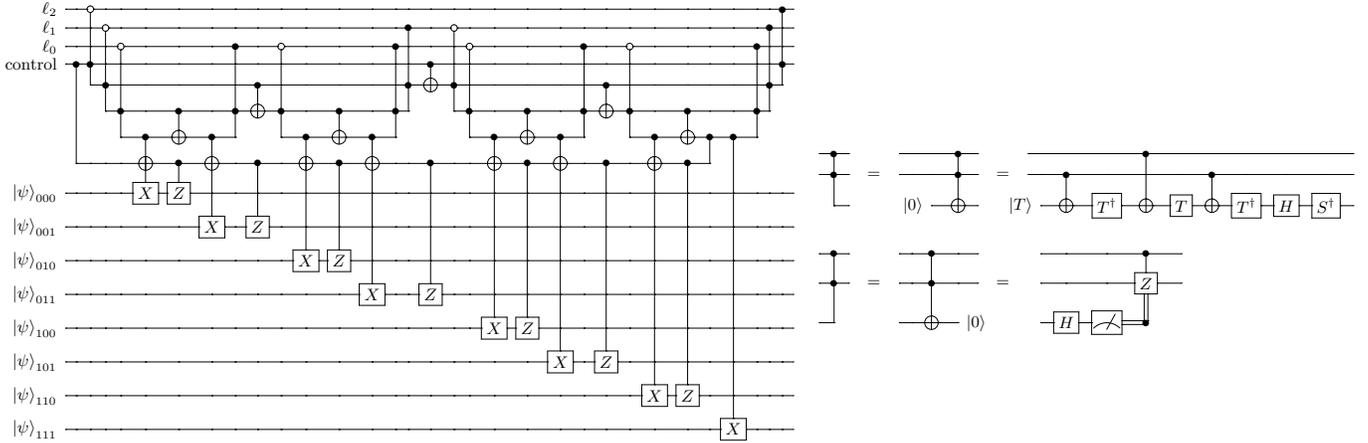

\subsection{The Hamiltonian Application Circuit}

The circuit referred to in \eq{select} as $U$ should act as
\begin{align}
\label{eq:select_syk}
U \ket{p}\ket{q}\ket{r}\ket{s} \ket{\psi} \mapsto \ket{p}\ket{q}\ket{r}\ket{s} \gamma_p \gamma_q \gamma_r \gamma_s \ket{\psi}.
\end{align}
To implement this on a quantum computer, the Majorana operators are represented by strings of Pauli operators according to the Jordan-Wigner representation.
We therefore need to apply
the following transformation
\begin{equation}
\label{eq:select_majorana}
\ket{\ell} \ket{\psi} \mapsto \ket{\ell} \gamma_\ell \ket{\psi} = \ket{\ell} X_{\ell} \cdot Z_{\ell-1} \cdot Z_{\ell -2} \cdots Z_{0} \ket{\psi}.
\end{equation}
Circuits for exactly this transformation were introduced in Section 3B of \cite{BabbushSpectra}, reproduced here in \fig{majorana}. The T complexity of this circuit implementing the transformation of \eq{select_majorana} is exactly $4N - 4$ and the Clifford complexity is also ${\cal O}(N)$. Clearly, four applications of this primitive are sufficient to implement $U$; thus, our total implementation has gate complexity $16 N - 16$ and uses only $\log N$ additional ancillae.

\section{Hamiltonian evolution}

Now we give more details on how to perform the Hamiltonian evolution and determine its cost.
A self-inverse operation $U$ is equivalent to a reflection.
Then $R$ provides a second reflection, and a product of reflections has a spectrum given by Theorem 1 of \cite{Szegedy2004} (a product of reflections gives a rotation in the same way as for Grover's algorithm \cite{Grover1996}).
The eigenvalues then correspond to the angle of rotation for each eigenstate.
Applying that theorem shows that ${\cal W}=R U$ has eigenvalues of $e^{\pm i \arccos(h/\lambda)}$ for eigenvalue $h$ of $H$ \cite{Low2016}. One can show that (see the proof of Lemma 6 of \cite{BerryFOCS15} or Lemma 16 of \cite{CKS17})
\begin{equation}
\label{eq:cheb}
\bra{G}{\cal W}^n\ket{G} = T_n(H/\lambda) ,
\end{equation}
where $T_n(\cdot)$ is the $n^{\rm th}$ Chebyshev polynomial of the first kind.
This result originates from the relation $T_n(\cos\theta)=\cos n\theta$. Here $\theta$ is equivalent to $h/\lambda$, and applying ${\cal W}^n$ gives a rotation by $n\theta$.
Using the Jacobi-Anger expansion
\begin{align}
\label{eq:anger}
e^{-iHt} = J_0(-\lambda t) + 2\sum_{n=0}^\infty i^n J_n(-\lambda t) T_n(H/\lambda) ,
\end{align}
where $J_n(\cdot)$ is the $n^{\rm th}$ Bessel function of the first kind, and we have used $\cos(\pm n \arccos(H/\lambda))=T_n(H/\lambda)$. 
Therefore a polynomial in ${\cal W}$ can be used to approximate $e^{-iHt}$, and this polynomial can be generated via quantum signal processing \cite{Low2017}.
The polynomial order needed to approximate $e^{-iHt}$ to within error $\epsilon$ is
\begin{equation}\label{eq:qspcost}
{\cal O}\left(\lambda t + \frac{\log (1/\epsilon)}{\log \log (1/\epsilon)}\right).
\end{equation}
For large $\lambda t$, it can be shown that the order required is approximately (see \app{jacobi})
\begin{equation}
\lambda t + \frac{3^{2/3}}2 (\lambda t)^{1/3} \log^{2/3}(1/\epsilon) .
\end{equation}
The complexity of simulating $e^{-iHt}$ via quantum signal processing is twice this, in terms of the number of applications of $R$ and $U$.
That can be seen from Theorem 1 of \cite{Low2016}, where the order is $N/2$, and the total number of controlled operations is $N$.

\section{Conclusion}

Using quantum signal processing, our simulation requires a number of applications of $A$, $B$, and $U$ given by \eq{qspcost}.
The total cost of simulation is thus, 
\begin{equation}
\label{eq:exact_scaling}
{\cal O}\left(\left(C_A + C_B + C_U\right)\left(\lambda t + \frac{\log (1/\epsilon)}{\log \log (1/\epsilon)}\right)\right)
\end{equation}
where $C_A$, $C_B$ and $C_U$ are the cost of implementing $A$, $B$ and $U$, respectively. In terms of gate complexities of the explicit implementations advocated for in this work, $C_A = {\cal O}({\rm polylog}(N/\epsilon))$, $C_B = {\cal O}(\log N)$, and $C_U = {\cal O}(N)$. We have found that $\lambda = {\cal O}(N^{5/2})$. Thus, the total asymptotic complexity becomes
\begin{align}
{\cal O}\left(N^{7/2} t + N^{5/2} t \,{\rm polylog}\left(N/\epsilon\right)\right).
\end{align}

Since we have compiled all bottleneck components down to Clifford + T gates, we are also able to report the leading order scaling of the T count of the algorithm as $\frac{2}{\sqrt{6}} N^{7/2} J t $. Assuming reasonable precision goals, for $N = 100$ the leading order T count is less than ten million $J t$ and for $N = 200$, the leading order T count is less than one hundred million $J t$. This should be compared to the roughly $10^{12}$ T gates required to simulate the active site of FeMoco with 108 qubits (a molecule relevant to Nitrogen fixation) \cite{Reiher2017,Motta2018,Berry2019}, $10^{9}$ T gates required to simulate 100-200 qubit interesting problems in solid-state electronic structure \cite{BabbushLow,BabbushSpectra} or the roughly $10^9$ T gates required to simulate a 100 qubit one-dimensional Heisenberg model for classically intractable durations \cite{Childs2017}. While the particular value of $J t$ would depend on the application, this analysis reveals that simulation of the SYK model is among the most viable applications of the first surface code quantum computers.\\

\subsection*{Acknowledgements} The authors thank Alexei Kitaev, Jarrod McClean, Zhang Jiang, Sergio Boixo, John Sous, Gavin Brennen, Yuval Sanders and Artur Scherer for
helpful discussions. Dominic Berry is funded by an Australian Research Council Discovery project (Grant No.
DP160102426).

\onecolumngrid
\bibliography{references,Berry}
\appendix

\section{Asymmetric Qubitization via Oblivious Amplitude Amplification}
\label{app:oaa}

An alternative approach to that presented in the main text is to use the fact that \begin{equation}
\label{eq:cheb}
\bra{G}{\cal W}^n\ket{G} = T_n(H/\lambda) ,
\end{equation}
holds for all steps in oblivious amplitude amplification, regardless of symmetry.
Define the step of oblivious amplitude amplification
\begin{equation}
\label{eq:walk}
{\cal U} = A^\dagger U^\dagger B R_0 B^\dagger U A,
\end{equation}
where $R_0 \equiv 2\proj{0} -\openone$ is a reflection on the control register.
This operation can very easily be made controlled, just by making the reflection $R_0$ controlled.
Then,
\begin{align}\label{even}
\bra{0}{\cal U}^n\ket{0} &= T_{2n}(H/\lambda) \\ \label{odd}
\bra{0}B^\dagger U A\,{\cal U}^n\ket{0} &= T_{2n+1}(H/\lambda),
\end{align}
Therefore, it is possible to generate both the even and odd terms required in \eq{anger} by using steps of ${\cal U}$ or steps of ${\cal U}$ followed by $B^\dagger U A$. In order to generate the polynomial required by quantum signal processing, an ancilla qubit can be used, which controls whether the odd or even terms in the polynomial in \eq{anger} are generated.
This qubit would primarily control the qubit rotations used in quantum signal processing.
Ultimately this qubit would also control the operations $B^\dagger U A$, applying them to produce the odd terms.

In this approach, combining the odd and even terms via a linear combination of unitaries gives a success amplitude that is not unity.
Amplitude amplification requires repeating the sequence of operations at least three times, giving an additional multiplicative overhead.
In the scheme proposed in the main text, $U$ is applied only once.
There is a factor of two for the preparations, but those have complexity that is logarithmic compared to the complexity of $U$.
Thus, the scheme in the main text is the most gate-efficient of the alternatives considered.

For completeness, we derive the equations in \eq{even} and \eq{odd}.
Let $P=\proj{G}$, and $R=2P-\openone$, and we require that $\bra{G}U\ket{G}=H$.
We claim that
\begin{equation}
\label{eq:odd}
U(RU^\dagger RU)^m P = (U-H)P U_{2m}(H)
+PT_{2m+1}(H)
\end{equation}
and
\begin{equation}
\label{eq:even}
(RU^\dagger RU)^m P = (H-U^\dagger)P U_{2m-1}(H)
+P T_{2m}(H).
\end{equation}
Let us consider \eq{odd} first.
For the case $m=0$ the right hand side is
\begin{equation}
(U-H)P + PH = UP = U(RU^\dagger RU)^0 P.
\end{equation}
Therefore this expression is clearly true for $m=0$.
Then, if it is true for $m-1$, we get
\begin{align}
U(RU^\dagger RU)^m P &= (URU^\dagger R) U(RU^\dagger RU)^{m-1} P\\
&=  (URU^\dagger R)  [(U-H)P U_{2m-2}(H)
+PT_{2m-1}(H)].\nonumber
\end{align}
Now
\begin{align}
(URU^\dagger R)  (U-H)P &= U(2P-\openone)U^\dagger(2P-\openone)UP - U(2P-\openone)U^\dagger(2P-\openone)HP
= U(2H^2-\openone)P - HP.
\end{align}
Also
\begin{align}
(URU^\dagger R) P &= U(2P-\openone)U^\dagger(2P-\openone) P 
= (2UH-\openone) P.
\end{align}
Therefore,
\begin{align}
\label{eq:almost}
U(RU^\dagger RU)^m P
&= [U(2H^2-\openone) - H]P U_{2m-2}(H) + (2UH-\openone)P T_{2m-1}(H) \\
&= U[(2H^2-\openone)U_{2m-2}(H) +2H T_{2m-1}(H)]P
-[HU_{2m-2}(H)+T_{2m-1}(H)]P.\nonumber
\end{align}
Using the standard relation $T_n(x)=[U_n(x)-U_{n-2}(x)]/2$ gives
\begin{align}
-[HU_{2m-2}(H)+T_{2m-1}(H)] &= 2HT_{2m}(H)-HU_{2m}(H)-T_{2m-1}(H) = -HU_{2m}(H) + T_{2m+1}(H)
\end{align}
where in the last line we have used the standard recurrence relation for $T$.
Next, using the recurrence relation $T_{n+1}(x) = xT_n(x)-(1-x^2)U_{n-1}(x)$ gives
\begin{align}
(2H^2-\openone)U_{2m-2}(H) +2H T_{2m-1}(H) &=2H T_{2m-1}(H)-2(\openone-H^2)U_{2m-2}(H) +U_{2m-2}(H)\\ & = 2T_{2m}(H) + U_{2m-2}(H) = U_{2m}(H)\nonumber
\end{align}
where in the last line we have used $T_n(x)=[U_n(x)-U_{n-2}(x)]/2$ again.
Therefore \eq{almost} gives
\begin{align}
U(RU^\dagger RU)^m P &=
U U_{2m}(H) P
 -HU_{2m}(H)P + T_{2m+1}(H)P \\
&= (U-H)P U_{2m}(H) + P T_{2m+1}(H) \nonumber
\end{align}
Hence \eq{odd} is true for all $m$ as required.

Next we look at \eq{even}.
We can start with \eq{odd} and apply $U^\dagger$, to give
\begin{align}
(RU^\dagger RU)^m P &= (\openone - U^\dagger H)P U_{2m}(H)+U^\dagger P T_{2m+1}(H) = P U_{2m}(H) + U^\dagger [-HU_{2m}(H) + T_{2m+1}(H)]P
\end{align}
Now we have
\begin{align}
-HU_{2m}(H) + T_{2m+1}(H) &= -U_{2m+1}(H) + 2T_{2m+1}(H) 
= -U_{2m-1}(H) 
\end{align}
where in the first line we have used $U_n(x)=xU_{n-1}(x)+T_n(x)$ and in the second line we have used $T_n(x)=[U_n(x)-U_{n-2}(x)]/2$ again.
Again using $U_n(x)=xU_{n-1}(x)+T_n(x)$ gives
\begin{equation}
U_{2m}(H) = HU_{2m-1}(H)+T_{tm}(H)
\end{equation}
Bringing these together, we have
\begin{align}
(RU^\dagger RU)^m P &= [HU_{2m-1}(H)+T_{2m}(H)]P - U^\dagger U_{2m-1}(H)P = (H-U^\dagger)P U_{2m-1}(H) + P T_{2m}(H)
\end{align}
This is \eq{even}, which we aimed to show.

Now note that, projecting onto $\ket{G}$ at the end, we have in the odd case
\begin{equation}
PU(RU^\dagger RU)^m P = PT_{2m+1}(H),
\end{equation}
which implies
\begin{equation}
\bra{G} U(RU^\dagger RU)^m \ket{G} = T_{2m+1}(H).
\end{equation}
In the even case we have
\begin{equation}
P(RU^\dagger RU)^m P = PT_{2m}(H),
\end{equation}
which implies
\begin{equation}
\bra{G} (RU^\dagger RU)^m \ket{G} = T_{2m}(H).
\end{equation}

\section{Scaling of Cutoff for Jacobi-Anger Expansion}
\label{app:jacobi}

For cutoff $K$ on the Jacobi-Anger expansion, the error is upper bounded by
\begin{equation}
\label{eq:anger}
\varepsilon \le 2\sum_{n=K+1}^\infty |J_n(-\lambda t)| = 2\sum_{n=K+1}^\infty |J_n(\lambda t)| .
\end{equation}
We will put $\tau=\lambda t$ for simplicity.
We are interested in the regime where the cutoff $K$ is large, and slightly larger than $\tau$.
This is known as the transition region, and the asymptotic form is given in Eq.~(10.19.8) of \cite{DLMF}.
The leading order term is
\begin{equation}
J_n(\tau) \sim \left( \frac 2n \right)^{1/3} {\rm Ai}(-2^{1/3} a),
\end{equation}
where $a=(\tau - n)/n^{1/3}$.
The Airy function has the asymptotic form (see \cite{DLMF} Eq.~(9.7.5))
\begin{equation}
{\rm Ai}(z) \sim \frac{e^{-(2/3)z^{3/2}}\Gamma(5/6)\Gamma(1/6)}{4 \pi^{3/2}z^{1/4}}.
\end{equation}
Using this asymptotic form gives, with $\delta=n-\tau$,
\begin{align}
J_{n}(\tau) &\sim \left( \frac 2n \right)^{1/3}\frac{e^{-(2/3)(\delta (2/n)^{1/3})^{3/2}}\Gamma(5/6)\Gamma(1/6)}{4 \pi^{3/2}(\delta (2/n)^{1/3})^{1/4}}
=
e^{-\frac{2\sqrt{2}\delta^{3/2}}{3\sqrt{n}}}\frac{\Gamma(5/6)\Gamma(1/6)}{2\time 2^{3/4}\pi^{3/2}(n\delta)^{1/4}} .
\end{align}
To leading order we can replace $n$ with $\tau$, giving
\begin{equation}
J_{n}(\tau) \sim e^{-\frac{2\sqrt{2}\delta^{3/2}}{3\sqrt{\tau}}}\frac{\Gamma(5/6)\Gamma(1/6)}{2\time 2^{3/4}\pi^{3/2}(\tau\delta)^{1/4}} .
\end{equation}
We can then approximate the upper bound on the error by integrating this asymptotic form
\begin{align}
\varepsilon &< 2\sum_{n=K+1}^\infty |J_n(\tau)| 
 \sim 2 \int_{\Delta}^\infty e^{-\frac{2\sqrt{2}\delta^{3/2}}{3\sqrt{\tau}}}\frac{\Gamma(5/6)\Gamma(1/6)}{2\time 2^{3/4}\pi^{3/2}(\tau\delta)^{1/4}} d\delta 
= \sqrt{\frac 23}{\rm erfc}\left( \frac{(2\Delta)^{3/4}}{\sqrt{3}\tau^{1/4}}\right),
\end{align}
where $\Delta=K-\tau$.
Now erfc has the asymptotic form (see Eq.~(7.12.1) of \cite{DLMF})
\begin{equation}
{\rm erfc}(z) \sim \frac 1{z\sqrt{\pi}} e^{-z^2}.
\end{equation}
We then get
\begin{equation}
\varepsilon \lesssim \sqrt{\frac 23}\frac{\sqrt{3} \tau^{1/4}}{\sqrt{\pi}(2\Delta)^{3/4}}
e^{-\frac{(2\Delta)^{3/2}}{2\tau^{1/2}}} .
\end{equation}
Solving for $\Delta$ gives
\begin{equation}
\Delta \lesssim  \frac {\tau^{1/3}}2 \left[ \frac 32 W\!\left( \frac 8{\varepsilon^2 \pi}\right) \right]^{2/3},
\end{equation}
where $W(z)$ is the Lambert W-function, which gives the principal solution of $w$ in $z=w e^w$.
For large argument $W(z)\sim \log(z)$ (see Eq.~(4.13.10) of \cite{DLMF}).
Therefore, for upper bound $\epsilon$ on the allowable error, we can take
\begin{equation}
K \sim \tau + \frac {3^{2/3}}2 \tau^{1/3} \log^{2/3}(1/\epsilon).
\end{equation}

\end{document}